\documentclass[10pt]{article}
\usepackage{amsfonts}
\usepackage{amsthm}

\usepackage{xspace,colortbl}
\usepackage[latin1]{inputenc}
\usepackage[T1]{fontenc}
\usepackage{graphicx}
\usepackage{graphics}
\usepackage{verbatim}
\usepackage{bookman}
\usepackage{amsmath}
\usepackage{amssymb,latexsym}
\usepackage{amscd}
\usepackage{mathrsfs}
\usepackage{pifont}
\usepackage{color}
\usepackage{graphicx}

\newcommand{\sss}{\setcounter{equation}{0}}
\newtheorem{theorem}{THEOREM}[section]

\newtheorem{question}[theorem]{OPEN PROBLEM}

\def\beq{\begin{equation}}
\def\ene{\end{equation}}

\begin{document}
\baselineskip=20 pt
\parskip 6 pt

\title{Hybrid resonance and long-time asymptotic of the solution to Maxwell's equations
\thanks{ PACS Classification (2010): 52.35.Hr; 52.25.Os; 52.40.Db; 52.55; 52.50.-b.  Mathematics Subject Classification(2010): 35Q61; 78A25. Research partially supported by projects  ANR under contract ANR-12-BS01-0006-01   and PAPIIT-DGAPA UNAM IN102215.  }}

\author{Bruno Despr\'es \\Laboratory Jacques Louis Lions, University
Pierre et Marie Curie, Paris VI,\\  Bo\^{\i}te courrier 187,
75252 Paris Cedex 05, France.\\ despres@ann.jussieu.fr 
 \and Ricardo Weder\thanks{Fellow Sistema Nacional de Investigadores}\\ Departamento de F\'{\i}sica  Matem\'atica, \\
Instituto de Investigaciones en Matem\'aticas 
Aplicadas y en Sistemas,\\
Universidad Nacional Aut\'onoma de M\'exico, Apartado Postal
 20-126, DF 01000, M\'exico.\\ weder@unam.mx }
%\date{}

\maketitle

\centerline{{\bf Abstract}}
%\begin{center}
\noindent We study the long-time asymptotic of the solutions to Maxwell's equation in the case of a {upper-}
  hybrid resonance in the cold plasma model. We base our analysis in the transfer to the time domain of the recent results of B. Despr\'es, L.M. Imbert-G\'erard and R. Weder,   J.  Math. Pures  Appl.  {\bf 101} ( 2014) 623-659, where the singular solutions to Maxwell's equations in the frequency domain  were constructed by means of a limiting absorption principle and a formula for the heating of the plasma in the limit of vanishing collision frequency was obtained.  Currently there is considerable interest in these problems, in particular, because upper-hybrid resonances are a possible scenario for the heating of plasmas, and since they can be a model for the diagnostics  involving  wave scattering in plasmas.

\noindent Corresponding author Ricardo Weder, Tel + 52 55 56223548, Fax + 52 55 56223596.

\section{Introduction}\sss

In plasma physics  \cite{brambila,chen,freidberg} upper-hybrid resonances may develop
at places where a density gradient of charged particles  excited by   a strong background magnetic field generates
singular solutions to Maxwell's equations. This phenomenon shows up   
in propagation of electromagnetic waves in the outer region of the  atmosphere, as explained first in 
\cite{budden}. It also appears   in reflectometry experiments \cite{gusakov,heuraux} and in heating devices
in fusion plasma \cite{dumont} in Tokamaks. An important  feature  in this direction is the  energy deposit which is finite. It   may exceed by far the energy exchange which occurs in Landau damping \cite{freidberg,villani}.
Notice however that there exists situations where hybrid resonance and Landau damping are modeled in a unique set 
\cite{peysson,peysson2}. {Furthermore, our model could be applied in diagnostics involving wave scattering at the upper-hybrid resonance \cite{gusakov2}, \cite{gusakov}.

The starting point of the analysis  is
 the linearized  Vlasov-Maxwell's equations
of a non-homogeneous plasma
around a bulk magnetic field $\mathbf B_0\neq 0$.
It yields the non-stationary Maxwell's equations with
a linear current
\begin{equation} \label{1.1}
\left\{
\begin{array}{ll}
-\frac1{c^2}\partial_t \mathbf{E}+
\nabla \wedge \mathbf{B }=\mu_0 \mathbf{J}, &
\mathbf{J}= -e N_e \mathbf{u}_e, \\
\partial_t \mathbf{B}+
\nabla \wedge \mathbf{E }=0, \\
m_e  \partial_t \mathbf{u}_e=
-e\left(\mathbf{E}+ \mathbf{u}_e \wedge \mathbf{B}_0   \right)-
m_e \nu  \mathbf{u}_e. 
\end{array}
\right.
\end{equation}
The electric field is $\mathbf E$ and the magnetic field is 
$\mathbf B$.
The modulus of 
the background magnetic field
$|\mathbf B_0|$ and its direction $\mathbf b_0=\frac{\mathbf B_0}{|\mathbf B_0|} $ will be  assumed  constant in space 
for simplicity in our work. The total magnetic field is expanded as first order as
$\mathbf B_{\rm tot}= \mathbf B_0+\mathbf B$. Note that in the last equation in \eqref{1.1} $\mathbf B$ is neglected.
The absolute value
of the charge of electrons is $e$, the mass of electrons is
$m_e$,
the velocity of light is $c=\sqrt{\frac1{\varepsilon_0 \mu_0}} $ where 
the permittivity of vacuum is $\varepsilon_0$ 
and  the permeability
of vacuum is $\mu_0$.
The third  equation  corresponds to moving electrons with velocity $ \mathbf{u}_e$,
and the  electronic density $N_e$
is a given function of the space variables.
One assumes the existence of { a bath of particles} which is the reason of the 
friction between the electrons and the {bath of particles}
with collision frequency $\nu$.
 Much more material about such models can be found in classical
physical textbooks \cite{brambila,freidberg}.
The loss of energy in domain $\Omega$
can easily be computed in the time domain
starting from
(\ref{1.1}).  One obtains
$$
\frac{d}{dt}
\int_\Omega \left(
\frac{\varepsilon_0 \left| \mathbf{E}\right|^2}{2}
+
\frac{ \left| \mathbf{B}\right|^2}{2\mu_0}
+ \frac{m_e N_e  \left| \mathbf{u}_e\right|^2}{2}
\right)=
-\int_\Omega {\nu m_e N_e    \left| \mathbf{u}_e\right|^2
}+
\mbox{ boundary terms}.
$$
Therefore,
\begin{equation}\label{1.2}
{\cal Q}(\nu)=\int_\Omega {\nu m_e N_e    \left| \mathbf{u}_e\right|^2
}
\end{equation}
represents the total loss of energy of the electromagnetic field
plus  the electrons in function of the collision frequency $\nu$. Since the 
 energy loss  is necessarily equal
to what is gained by the bath of particles, it will be referred
to as the heating.  In fusion plasma, a value of $\nu\approx 10^{-7}$ in relative units is often encountered.
It is therefore tempting to set the friction parameter, i.e. the collision frequency $\nu$ ,  equal to zero, but this  naive approach is incorrect, as we explain below.

Equations   \eqref{1.1}  can be written in the frequency domain, where $\omega$ is the frequency, that is  
$\partial_t  = -i\omega$ where for simplicity of the notations
$\omega>0$.  We assume that the bulk magnetic field is along the $z$ coordinate, i.e. $\mathbf b_0= (0,0,1).$
We obtain,
\begin{equation}\label{1.3}
\left\{
\begin{array}{ll}
\frac1{c^2}i\omega  \mathbf{E}+
\nabla \wedge \mathbf{B }=-\mu_0 e  N_e \mathbf{u}_e, \\
-i\omega  \mathbf{B}+
\nabla \wedge \mathbf{E }=0, \\
-im_e  \omega \mathbf{u}_e=
-e\left(\mathbf{E}+ \mathbf{u}_e \wedge \mathbf{B}_0   \right)-
m_e \nu  \mathbf{u}_e. 
\end{array}
\right.
\end{equation}
One can
 compute the velocity using the  third equation $\widetilde{\omega} \mathbf{u}_e+
\omega_c i  \mathbf{u}_e \wedge \mathbf{b}_0  
= -\frac{e}{m_e}i\mathbf{E}$  
 where the cyclotron frequency is $\omega_c= \displaystyle\frac{e|\mathbf{B}_0|}{m_e}$.
  The frequency 
$
\widetilde{\omega}=\omega +i \nu$ is shifted in the complex plane
by a factor equal to the friction parameter. It is then easy to eliminate $\mathbf{u}_e$ from the first equation
of the system (\ref{1.3})
and to obtain the time harmonic Maxwell's equation
\begin{equation} \label{1.4}
\nabla \wedge  \nabla \wedge
 \mathbf E - \left( \frac{\omega}{c} \right)^2 
\underline{\underline{\varepsilon}}(\nu)
\mathbf  E = 0.
\end{equation}
The dielectric tensor  is the one of the cold plasma approximation, the so-called Stix tensor,  
\cite{chen,freidberg}
\begin{equation} \label{1.5}
\underline{\underline{\varepsilon}}(\nu)=
\begin{pmatrix}
1-\frac{\widetilde{\omega}\omega_p^2}{
\omega
\left(\widetilde\omega^2-\omega_c^2\right)}&i \frac{\omega_c \omega_p^2}{\omega \left(\widetilde\omega^2-\omega_c^2\right)}&0 \\
-i \frac{\omega_c \omega_p^2}{\omega \left(\widetilde\omega^2-\omega_c^2\right)}
&1-\frac{\widetilde{\omega} \omega_p^2}{
\omega\left( \widetilde\omega^2-\omega_c^2\right) }&0 \\
0&0& 1 - \frac{\omega_p^2}{\omega \widetilde{\omega}}
\end{pmatrix}
.\end{equation}
The parameters of the dielectric tensor are
the cyclotron frequency 
$
\omega_c
$
 and 
the plasma frequency
$
\omega_p=\sqrt{\frac{e^2 N_e}{\varepsilon_0m_e}  }
$
 which depends on the electronic density $N_e$. 
 We are interested  in  the physical situation where the electronic density   $N_e$ is not constant, that is $\nabla N_e\neq 0$.
Observe that the cyclotron frequency $\omega=\omega_c$ is a singularity of the dielectric tensor. In this paper we consider $\omega\neq \omega_c$, hence, the dielectric tensor \eqref{1.5} is  smooth.
 In fact, we  focus on the paradoxical {upper-hybrid }resonance that appears when   $\omega= \omega_h:= \sqrt{\omega_c^2+\omega_p^2}$.  In plasma physics $\omega_h$ is called the upper-hybrid frequency.
 
 If we set  $\nu=0$ the first two diagonal entries in the Stix  tensor are  equal to $\frac{\omega^2- \omega_H^2}{\omega^2-\omega_p^2}$. The crucial issue is that they are equal to zero when $ \omega=\omega_H$ and that they continuously change in sign when $\omega$ increases from values smaller than $\omega_H$ to values bigger than $\omega_H$.   For this reason (see \cite{jmpa}) the system   \eqref{1.4} with $\nu=0$ is ill posed: the solution is not unique and, furthermore, it has singular solutions that contain distributions. The way out of this dilemma  \cite{jmpa} is a limiting absorption principle,  we take $\nu$  to zero in a limiting sense: We consider $ \nu >0$ small and we construct an unique solution to  \eqref{1.4} characterized by its behaviour at the point in space where $ \omega_H=\omega$ and by demanding that it goes to zero at spacial infinity, away from the sources of the electromagnetic field (the system \eqref{1.3} is assumed to be coercive (non propagating) at infinity). We call it the singular solution. We then prove that as $ \nu \downarrow 0$ the singular solution converges (in distribution  sense) to a limiting singular solution that is the appropriate physical solution to our problem. Furthermore, we give a formula for the limiting heating $\mathcal Q(0^+):= \lim_{\nu \downarrow 0} \mathcal Q(\nu)$, that turns out to be positive.  The fact  $Q(0^+)>0$  implies that the hybrid resonance is able to transfer a finite amount of energy from the electromagnetic field and the electrons to the bath of particles ( i.e. to heat the bath of particles) even in the limit  when $ \nu = 0$. This is, indeed, a remarkable result.  The  physical interpretation is that  as $ \nu$ goes to zero -and the solution becomes singular- the velocity of the electrons increases (since the friction with the ions goes to zero) and there is a compensation in  the right-hand side of \eqref{1.2}. In the end, the increase in the velocity of the electrons dominates,  so that, in the limit the heating $\mathcal Q (0^+)$ is positive. Moreover, from the mathematical point of view, the fact that the singular solution with $\nu >0$ converges to a limiting singular solution implies  that for $\nu$ small the singular solution is close to the limiting singular solution and, hence,   it does not change very much with $\nu$.
 This means that a small $\nu$ positive can be used as a regularization parameter to numerically compute the limiting singular solution and the limiting heating $\mathcal Q (0^+)$. This numerical scheme has been successfully used in the  Ph. D. thesis 
\cite{lm} that contains extensive numerical calculations. It was found that the numerical solution with small $ \nu >0$ converges fast to the exact solution (with $\nu=0$) in point-wise sense, except of course, at the singularity. Moreover, a large fraction of the energy of the incoming wave may be absorbed by the bath of particles,  up to  $95 \%$ in the case of normal incidence, and up to $76.7 \%$ in the case of  oblique incidence. Our results  in  \cite{jmpa}, \cite{lm}, in particular our formula for the heating $\mathcal Q(0^+)$ shows, in a rigorous and quantitative way, that upper-hybrid resonances are, indeed, an efficient method to heat the bath of particles.    

We now present, for later use,  our results in \cite{jmpa} in a precise way.  To study the upper-hybrid resonance we consider
the $2\times 2$ upper-left block in \eqref{1.4}, that corresponds to the transverse electric (TE) mode, $E=(E_x,E_y,0),$ where the electric field is transverse to the bulk magnetic field $\mathbf B_0$.  We  assume that we have a  slab geometry:
all coefficients in \eqref{1.5} depend only on the coordinate $x$. Furthermore, we suppose that  that $E_x,E_y$, are independent of  $z$, that is the coordinate along the bulk magnetic field $\mathbf B_0$.

Then, in the limit case $\nu=0$, the $2\times 2$ upper-left block in \eqref{1.4} gives
\begin{equation}\label{1.6}
\left\{\begin{array}{rrrr}
W &+\partial_y E_x &- \partial_x E_y & =0,
\\  \partial_y W &-\alpha  E_x &-
i\delta  E_y&  =0,
\\ -\partial_x W &+ i\delta
E_x 
&-\alpha  E_y &=0,
\end{array} \right. 
\end{equation}
where we find convenient to introduce the vorticity  $W: =  \partial_x E_y-   \partial_y E_x $ that is  proportional to the    magnetic field $B_z$.
 The coefficients $\alpha, \delta$ are equal to
\begin{equation} \label{1.7}
\alpha= \frac{\omega^2}{c^2} \left(
1-\frac{\omega_p^2}{\omega^2-\omega_c^2}
\right)
\qquad
\mbox{ and }
\qquad
\delta=
 \frac{\omega^2}{c^2} \times 
\frac{\omega_c \omega_p^2}{\omega \left(\omega^2-\omega_c^2\right)}.
\end{equation}
In  plasma physics  the  system \ref{1.6}
is called   the equations for the X-mode,   or the extraordinary mode, and also the extraordinary wave.

We  consider the  system   \eqref{1.6}  in the two dimensional domain, 
$$
\Omega=\left\{(x,y)\in \mathbb R^2, \quad
-L\leq x, \quad
y\in \mathbb{R}, \quad L >0\right\}.
$$
We assume the    non-homogeneous 
 boundary condition 
\begin{equation} \label{1.8}
 W+ i\lambda n_x E_y=g \mbox{ on the left boundary }x=-L,
\qquad \lambda>0,
\end{equation}
that corresponds to a source, like a radiation  antenna that is used to heat the plasma..
We suppose that $\alpha$ and $\delta$ satisfy conditions that correspond to a upper-hybrid resonance at $x=0$. The main assumptions are: The function $\alpha$  is twice continuous differentiable and $ \delta$ is continuous with continuous first derivative. Furthermore, $ \alpha(0)=0, \alpha'(0) <0$, and $ \alpha \neq 0$ for $ x \neq 0$. Furthermore, $ \delta(0)\neq 0$.  We assume  that the coefficients
are constant far enough from the source, i.e.  : that there exist    $\delta_\infty$, 
$\alpha_\infty$, and $ H>0$   such  that
$$
\delta(x)=\delta_\infty \mbox{ and }\alpha(x)=\alpha_\infty,\qquad 
H\leq x < \infty,
$$
and that our system is coercive  (non propagative) at infinity,
\begin{equation}\label{1.9}
\alpha_\infty^2-\delta^2_\infty>0.
\ene
In  \cite{jmpa} we also assume other technical conditions. For a complete list   see  assumptions $H_1- H_6$  in \cite{jmpa}.

{For the purpose of considering the limiting absorption principle we actually consider a  model where the Stix tensor  \eqref{1.5}  is replaced by  
$ \underline{\underline{\varepsilon}}(\nu)=\underline{\underline{\varepsilon}}(0)+ i \nu I$, where, of course, $\underline{\underline{\varepsilon}}(0)$ is the value of the Stix tensor \eqref{1.5} for $ \nu=0$. This is a simplified linear approximation. For the two by two block in \eqref{1.5}, that corresponds to the upper-hybrid resonance, this amounts to replace $\alpha$ by $\alpha+ i \nu$ in \eqref{1.6}. We denote the solution by, $\left( E^\nu_x, E^\nu_y, W^\nu\right)$. Note that when $\nu >0$ our system  \eqref{1.7} is well posed. Clearly, $\nu$ in  our  linear model is not exactly the same quantity as the collision frequency $\nu$ in (\ref{1.1}, \ref{1.2}, \ref{1.3}, \ref{1.4}). However, since it plays the same role, we still denote it by $\nu$ and we  call it the collision frequency, to simplify the notation. }  Furthermore,  we take advantage of the fact that the coefficients $\alpha, \delta$ only depend on $x$ to take a Fourier transform along $y$,
$$
\mathbf U^{\theta,\nu}=
\left(
\begin{array}{c}
U^{\theta,\nu} \\
V^{\theta,\nu} \\
W^{\theta,\nu}
\end{array}
\right):= \int_{\mathbb R}\, e^{-i\theta y}\, 
\left(\begin{array}{c}
E_x^\nu \\
E_y^\nu \\
W^\nu
\end{array}
\right)\, dy.
$$
Then, $\mathbf U^{\theta,\nu}$ is a solution to the system
\begin{equation}\label{1.10}
\left\{
\begin{array}{l}
  W^{\theta,\nu} 
+  i\theta U^{\theta,\nu}  - \frac{d}{dx} V^{\theta,\nu}  =0, \\
  i\theta W^{\theta,\nu}  -(\alpha (x)+i\nu) U^{\theta,\nu} 
-i\delta (x) V^{\theta,\nu} =0,\\ 
  - \frac{d}{dx} W^{\theta,\nu}  
+ i\delta (x) U^{\theta,\nu} 
-(\alpha(x)+i\nu) V^{\theta,\nu}  =0.
\end{array}
\right.
\end{equation} 
 We  denote by 
$$
\mathbf{U}^{\theta,\nu}=
(U^{\theta,\nu},V^{\theta,\nu},W^{\theta,\nu})
$$
the solution to \eqref{1.10}  that  is uniquely defined by the following two conditions \cite{jmpa}:
\begin{enumerate}
\item  It is  exponentially decreasing at infinity. Recall that by \eqref{1.9}  the system \eqref{1.10} is coercive (non propagative) for $ x \geq H$. 
\item 
Its value at the origin is  normalized with the requirement
\begin{equation} \label{1.11}
i\nu U^{\theta,\nu}(0)=1.
\end{equation}
\end{enumerate}
Note that as $ \nu \downarrow 0$ the value of $ U^{\theta,\nu}(0)$ goes to infinity and for this reason we call it the singular solution.  Observe  that in \cite{jmpa} we denote this solution by  $\mathbf{U}_2^{\theta,\nu}$ to distinguish  it 
from other solutions  introduced there.

In Proposition 5.16 of \cite{jmpa} we proved that,

\beq\label{1.12}
\lim_{\nu \downarrow 0}\mathbf{U}^{\theta,\nu}=
\mathbf{U}^{\theta,+}:=
\left(
P.V. \frac1{\alpha(x)} + \frac{i \pi}{\alpha'(0)} \delta_D
+ u^{\theta,+}, \;  
v^{\theta,+}, \; w_2^{\theta,+}
\right)
\ene
where $u^{\theta,+},
v^{\theta,+}, w^{\theta,+}\in L^2(-L,\infty)$
and $\delta_D$ is the Dirac delta function  at the origin.

The limit $\mathbf{U}^{\theta,+}$
is a solution in the sense of distributions  of the system (\ref{1.10})  with $\nu=0$. We call it  the limiting  singular solution.
Furthermore, in Remark 10 of \cite{jmpa} we prove that the functions $u_2^{\theta,+},
v_2^{\theta,+}, w_2^{\theta,+}$ can be obtained as the unique, properly normalized,  solution  of the  system of equations.

$$
\left\{
\begin{array}{clc}
w_2^{\theta,+}-\frac{d}{dx}v^{\theta,+}_2   +i \theta u_2^{\theta,+}
  &=&-i \theta P.V.  \frac{1}{\alpha(x)} + \frac{\theta \pi}{\alpha'(0)}
 \delta_D,
  \\
i\theta  w_2^{\theta,+} -\alpha(x)  u^{\theta,+}_2 
 -i \delta(x)   v^{\theta,+}_2  &=
&1,
 \\
-\frac{d}{dx} w^{\theta,+}_2 +i \delta(x)
u^{\theta,+}_2    - \alpha(x) v^{\theta,+}_2 
&=&- iP.V.  \frac{\delta(x)}{\alpha(x)}+  \frac{\delta(0) \pi}{\alpha'(0)} \delta_D.
\end{array}
\right.
$$
There is a striking similarity with scattering theory {  \cite{yaf}},  where  the solutions obtained by the
 limiting absorption principle are characterized as the unique solutions that
satisfy the radiation condition, that is to say,  they  are uniquely determined by the
behavior at infinity. Here, in order that the  singular solutions are uniquely determined we have to specify  
 their behavior at $+\infty$ and also  their singular part at $x=0$.
$
P.V. \frac1{\alpha(x)} + \frac{i \pi}{\alpha'(0)} \delta_D.
$
From the mathematical point of view it is natural that we have to specify the singular part because our  equations are degenerate at $x = 0$. The physical interpretation of this fact is that $x=0$ acts as a scattering channel,  and then, a condition has to be specified at $x=0$. {Similar situations arise in scattering theory, in the case   of obstacles with a singular boundary \cite{hempel1} and for manifolds with ends \cite{hempel2}.   }

 We construct the physical singular solution  (in distribution sense) to \eqref{1.6} by inverting the Fourier transform along $y$ and by demanding that the boundary condition \eqref{1.8} be satisfied.  Let us designate  by $\widehat{g}$ the Fourier transform of  the function $g$  that appears in the boundary condition \eqref{1.8},
$$
\widehat{g}(\theta):=  \int_{\mathbb R}\,g(y)\, e^{-i\theta y}\, dy.
$$

In Theorem 1.1 of \cite{jmpa}  we prove the following results: If  $ g \in L^2(\mathbb R)$ with $ \widehat{g}$ of compact support, 
there exists a    solution  (in distribution sense)
of (\ref{1.6}) with boundary condition (\ref{1.8}) that goes to zero infinity.
This solution, that we call the physical singular solution, has the following  representation formula 
\begin{equation} \label{1.13}
\left(
\begin{array}{c}
E_x^+ \\
E_y^+ \\
W^+
\end{array}
\right)
(x,y)=\frac1{2\pi}
\int_{\mathbb{R}}
\frac{ \widehat{g}(\theta)}{   \tau^{\theta,+}}\,  \mathbf{U}^{\theta,+}  \, e^{i\theta y  }d\theta.
\end{equation} 
The transfer coefficient
$ \tau^{\theta,+}$ is given by,
$$
\tau^{\theta, +} := W^{\theta,+}(-L)+i \lambda V^{\theta,+}(-L).
$$
 The electric field $E_x$
does not belong to $L_{\rm Loc}^1\left((-L,\infty)\times  \mathbb R\right)$ (of course, unless the source  $g$ is identically zero). The other components  are always more regular: In particular
  $E_y^+,W^+\in L^2\left((-L,\infty)\times  \mathbb R\right)$, but they  have a logarithmic singularity at $x=0$, see equations (5.57), (5.58) in \cite{jmpa}.

Moreover, the value of the resonant 
heating is
\beq \label{1.15}
{\mathcal Q}(0^+): = \lim_{\nu \downarrow 0} {\mathcal Q}(\nu)=
\frac{\omega\, \varepsilon_0}{2}\,
\int_{\mathbb{R}}  \frac{  \left|
\widehat{g}(\theta) \right|^2 }{ \left|
\alpha'(0)
\right|
 \left|  \tau^{\theta,+} \right|^2 }
d\theta>0.
\ene

Note that it is known since the seminal work of Budden \cite{budden}, \cite{chen} that the solution to the X-mode equations (in  a particular case that has explicit solution )  have a singularity as $1/x$ at the position (x=0) of the upper-hybrid resonance.
In the recent paper  \cite{lalide} the singularity around the hybrid resonance has been studied with  a purely local method based on deformation in the complex plane. Our results and our method in \cite{jmpa} are fundamentally different. Note that the fact that a singularity as $1/x$ ( that is not integrable in a neighborhood of zero ) appears in the solution means that  it is not a solution in distribution sense. Of course, the term $1/x$ can be interpreted as a principal value, but also a Dirac delta function, or another distribution with support at $x=0$ could be added. Our global analysis in \cite{jmpa}, based in the limiting absorption principle and the requirement that the solution goes to zero at infinity, is the crucial fact that allowed us to calculate the singularity of the physical solution given in \eqref{1.13}. Furthermore, this was essential to compute the heating   \eqref{1.15} that depends on the singularity of the physical solution. To our knowledge our paper \cite{jmpa} is the first that  gives a formula for the  heating of an upper-hybrid resonance with a discussion to the radiation condition at infinity.

   The results of \cite{jmpa} pose challenging new mathematical and physical questions, for example:
   
   \begin{question}{\rm
    To study the propagation, in the time domain, of a wave packet 
   (a definition in the context of waves in plasmas is to be found in  \cite{brambila}, see also
   \cite{lighthill,stix}) that is incident from the left on   the upper-hybrid resonance.
   
   Our results in \cite{jmpa} in the frequency domain make it intuitively clear to expect that part of the incoming wave packet will be reflected and that the transmitted part will be absorbed by the  bath of particles  as heating. To     set-up this  translation  of our results from the  frequency domain  to the time-domain is an issue that has to be treated with mathematical care because of the singularity of the physical solution.  In Section 2 we solve this problem. We prove that as $ t \rightarrow \pm \infty$  the   part of the wave packet   in $[-R,-L]$ tends to zero for every $ R>0$. Since there is no heating for $ x \leq -L$ this implies that the reflected part of the wave packet actually travels to $-\infty$ in the $x$ direction. Furthermore, we prove that as  $ t \rightarrow \pm \infty$ the wave packet tends to zero in the domain $ x \geq L$. Since we  assume that our problem is coercive (non propagative)  for $ x \geq H$ the wave packet can not travel to plus infinite in the $x$ direction. This implies that for large times  the transmitted part of the wave packet is absorbed by the bath of particles as heating.}
\end{question}

 \begin{question}{\rm
  Determine a  general perturbation theory and appropriate  limiting absorption  and limit amplitude principles  for the Maxwell equations   with hermitian  complex valued continuous dielectric tensors $\underline{ \underline \varepsilon } (\mathbf x)= 
 \underline{ \underline \varepsilon } (\mathbf x)^*$ with continuous eigenvalues which change sign.

 Here perturbation is for the dielectric tensor. For example one may use
 $\underline{ \underline \varepsilon }_\nu=\underline{ \underline \varepsilon } +\nu \mathbf I_3$ with
 $\mathbf I_3$ the identity matrix in $\mathbb C^3$, or more generally, as in the case of the   the cold plasma dielectric tensor (\ref{1.5}) which admits the expansion 
 $\underline{ \underline \varepsilon }_\nu=\underline{ \underline \varepsilon }_0+i \mathbf D +O(\nu^2)$
 with $\mathbf D$ a physically based hermitian non negative matrix that depends linearly on $\nu$   \cite{lalide}. Proving a (generalized) limiting absorption principle in these situations is an important and challenging issue.  
 In this direction some inspiration can perhaps
be obtained by comparison with   other non-standard dielectric tensors
and related physical and mathematical problems that can be found in
 \cite{bordeaux,bonnet1,bonnet2,chen:lipton,weder2,weder3}.
 
 Roughly speaking, the limit amplitude principle means that in the case of Maxwell equations in the time domain  with a periodic in time force term, the solutions  oscillate with the frequency of the force for large times. For standard dielectric tensors the limit amplitude principle follows from the limiting absorption principle    \cite{dautray:lions, leis,weder}. It is important to extend the validity of the limit amplitude principle to non-standard dielectric tensors as above.
 
 These are, indeed, challenging problems whose solution does not follow from the standard results  in  \cite{baba,cessenat,dautray:lions,leis,monk, weder}. }
 \end{question}

\section{Large time asymptotic of the solution to Maxwell's equations}\sss

To study the propagation of a wave packet we consider the problem in the whole space. We suppose that the assumptions that we made in the Introduction  and in \cite{jmpa} are valid for $ x \geq -L$, and furthermore, that $N_e(x)=0$ for $   x \leq -L$.  Then,    for $ - \infty <x \leq -L$  we have an isotropic and homogeneous medium (vacuum) with permittivity  $\varepsilon_0$ and  susceptibility $ \mu_0$. Hence, for $ x \leq -L$  and after taking a Fourier transform in the $y$ variable we see that   $\mathbf U^{\theta,+}$ must be  a solution to the system,

\begin{equation}\label{2.1}
\left\{
\begin{array}{l}
  W^{\theta,+} 
+  i\theta U^{\theta,+}  - \frac{d}{dx} V^{\theta,+}  =0, \\
  i\theta W^{\theta,+}  - \frac{\omega^2}{c^2} U^{\theta,+} 
 =0,\\ 
\frac{d}{dx}W^{\theta,+} + \frac{\omega^2}{c^2} V^{\theta,+}  =0,
\end{array}
\right.
\end{equation} 
for $ x \leq - L$.

If we eliminate   $  U^{\theta,+} $
in (\ref{2.1}) we obtain the following system of 
 ordinary differential  equations
\begin{equation} \label{2.2}
\frac{d}{dx}
\left(
\begin{array}{c}
 V^{\theta,+} \\
 W^{\theta,+}
\end{array}
\right)=
A^{\theta}(x)
\left(
\begin{array}{c}
 V^{\theta,+} \\
 W^{\theta,+}
\end{array}
\right)
\end{equation}
 where,
\begin{equation} \label{2.3}
{ A^{\theta}(x) }=
{
\left(
\begin{array}{cc}
0 &
1-\frac{\theta^2\, c^2}{ \omega^2}
\\
-
\frac{\omega^2}{c^2}
 &
0\end{array}
\right)}.
\end{equation}

We define $\left( V^{\theta,+},
 W^{\theta,+}
   \right)$  for $ x \leq -L$ as the unique solution of \eqref{2.2} with initial values at $x=-L$  given by the values, $\left( V^{\theta,+},
 W^{\theta,+}
   \right),$  of the singular  solution \eqref{1.12} at $x=-L$. Furthermore, we define $ U^{\theta,+}$ for $x \leq -L$ by the second equation in \eqref{2.1},  $ U^{\theta,+}=\frac{i \theta c^2}{\omega^2}   W^{\theta,+}$. In this way the singular solution ${\mathbf U}^{\theta, +}$  is defined for all $ x \in {\mathbb R}$.
   
   Assuming that $ \omega > c |\theta |$ the system \eqref{2.1} has two independent plane-wave solutions for $ x < - L$
   \beq\label{2.4}
    \left( V^{\theta }_\pm,
 W^{\theta,}_\pm
   \right) := \left( 1, \pm i \frac{\omega^2}{c^2 k(\omega)}   \right)\, e^{\pm i k(\omega)\, x}, \qquad k(\omega):= \sqrt{\frac{\omega^2}{c^2}- \theta^2}.
\ene
Note that the condition $ \omega > c |\theta |$ implies $k(\omega)\in \mathbb R$ which yields truly propagative plane-waves.
Defining  $ U^{\theta}_\pm=\frac{i \theta c^2}{\omega^2}   W^{\theta,}_\pm $, we obtain a basis of plane-wave  solutions to \eqref{2.1} for $ x \leq -L$,

\beq \label{2.5}
{\mathbf Z}^{\theta}_\pm= \left(  U^{\theta}_\pm, V^{\theta,}_\pm, U^{\theta }_\pm \right).
\ene
Then, for some functions $I(\omega), R(\omega)$ 
\beq \label{2.6}
{\mathbf U}^{\theta, +}= I(\omega) \, {\mathbf Z}^{\theta}_+ + R(\omega) \, {\mathbf Z}^{\theta}_-, \qquad \mathrm{for }\ \, x \leq -L.
 \ene
The functions $I(\omega), R(\omega)$ can be easily computed in terms of the values at $ x=- L$ of of ${\mathbf U}^{\theta,+}$ in \eqref{1.12}, we omit the details.

To consider a wave packet with frequency support in an interval $[\omega_0, \omega_1$ we make the following  assumptions  to make sure that the upper-hybrid resonance  takes place at a unique point $x_\omega$ for all   $ \omega \in [\omega_0, \omega_1]$. We have that

 $$
\alpha := \frac{\omega^2 (\omega+\omega_h)}{ c^2 (\omega^2 - \omega_c^2)}\, (\omega-\omega_h).
$$ 
  Recall that $\omega_c:= \frac{e|B_0|}{m_e}$, $ \omega_p:= \sqrt{\frac{e^2 N_e}{\varepsilon_0 m_e}}$ and that
  $\omega_h:=  \sqrt{\omega_c^2 + \omega_p^2}$ is the upper-hybrid frequency. As above,  assume that the density of electrons $N_e$ is zero  for $x \leq -L$,  that it is strictly increasing   for  $ -L \leq  x \leq H,$  that it is constant for $ x \geq H$ and that it is twice continuously differentiable.  We take $ \omega_0 > \omega_c$ and $\omega_1 < \omega_h(H)$. Then,  for $ \omega \in [\omega_0, \omega_1] $,  $ \alpha_\omega(-L)>0$ and $ \alpha_\omega(H) < 0$ with a unique point  $ x_\omega \in (-L, H)$ where $ \alpha_\omega (x_\omega)=0$ and ${\alpha}'(x_\omega):= \frac{d}{d x}  \alpha_\omega(x_\omega)  <0$. 

We apply the results in Section 1 and \cite{jmpa} for each $\omega \in [\omega_0, \omega_1]$ by translating $x_\omega$ to zero and we make explicit the dependence in $\omega$ of the singular solution denoting it by $  {\mathbf U}^{\theta,+}_\omega. $
We assume that $ \omega_0 > c|\theta|$ and we consider  a wave packet that is incident from the left,

\beq \label{2.7}
P_\varphi(t):= \int_{\mathbb R }\, e^{-i \omega t}\, A(\omega) \left( {\mathbf U}^{\theta,+}_\omega, \varphi \right) d\omega,
\ene
where $A(\omega)$ is an integrable  function with support contained in $ [\omega_0, \omega_1]$,   (this means that  $A(\omega)$  can be  different from  zero only in  $ [\omega_0, \omega_1]$ ). The function $ \varphi$  is continuously differentiable  on $ \mathbb R$,  with  values in $\mathbf C^3$, and has compact (closed and bounded) support. By $\left( \cdot,\cdot   \right)$
we denote the duality between the space of distributions and the test functions,

$$
 \left( {\mathbf U}^{\theta,+}, \varphi \right)= \int_{\mathbb R}\, {\mathbf U}^{\theta,+}_\omega(x)\, \varphi (x)\, dx.
$$
Note that by \eqref{2.6} $A(\omega) \, I(\omega)$ is the amplitude of the plane wave that is incoming from $- \infty$ and that  $A(\omega) \, R(\omega)$ is the amplitude of the reflected plane wave. Furthermore, as  ${\mathbf U}^{\theta,+}$ is a distribution it does not make sense to evaluate it in point-wise sense nor to compute its $L^2$ norm. This is the reason why we introduced in \eqref{2.7} the test function $\varphi$.  As $\varphi $ has support in a  compact set, say $K$, in intuitive terms by testing the wave packet with $\varphi$ we obtain information in how the wave packet behaves in $K$ as a function of time. Note that, in a naive way we could try to define,      
\beq \label{2.8}
 \int_{\mathbb R }\, e^{-i \omega t}\, A(\omega)  {\mathbf U}_\omega^{\theta,+}(x)\, dx,
\ene
but, of course, since ${\mathbf U}^{\theta,+}$ is a distribution this integral has to be given a meaning in  mathematical sense. 
The quantity  $P_\varphi(t)$ in \eqref{2.7} can be understood as a proper definition (or as as a regularized version) of \eqref{2.8}, where by changing the function $ \varphi$ we test the wave packet in different compact set in $ \mathbb R$. Actually as ${\mathbf U}_\omega^{\theta,+}$ is a function away from the hybrid resonance at $x=0$,  we could study  the behavior of the wave packet in point-wise sense, or in local $L^2$ norms, away from the upper-hybrid resonance, but this is not the point, because we wish  to consider precisely the influence of the upper-hybrid resonance, and hence we do not dwell on this issue here. 

We will use the Riemann-Lebesgue theorem. For the reader's convenience we recall it here. Let $\psi$ be a continuously differentiable   function defined on $\mathbb R$, with values on $\mathbb C^3$ and with compact support. Let $\widehat{\psi}$ be its Fourier transform,
$$
\widehat{\psi}(t):= \int_{\mathbb R}\, e^{-i t x}\, \psi(x)\, dx.
$$
Integrating by parts one proves that for some constant $C$,
$$
|\widehat{\psi}(t)| \leq C\, \frac{1}{|t|}, \qquad t \in \mathbb R.
$$
Furthermore, if $ \psi$ is integrable, approximating it in the  $L^1$ norm  by continuously differentiable functions with compact support 
one  proves that,
\beq\label{2.9}
\lim_{t \rightarrow  \pm \infty}\, \widehat{\psi}(t) =0.
\ene
Let us  decompose ${\mathbf U}^{\theta,+}_\omega$ between a regular part 
${\mathbf U}^{\theta, +}_{\omega, 1}$ and part ${\mathbf U}^{\theta, +}_{\omega,2}$ that contains all the singularity
  \beq \label{2.10}
  {\mathbf U}^{\theta ,+}_{\omega}= {\mathbf U}^{\theta, +}_{\omega, 1}+{\mathbf U}^{\theta, +}_{\omega,2},
  \ene
  where, denoting by $ \chi_O(x)$ the characteristic function of the set $O$, i. e. , $\chi_O(x)=1, x \in O, \chi_O(x)=0, x \notin O,$  
  \beq\label{2.11}
  {\mathbf U}^{\theta,+}_{\omega,1}:= {\mathbf U}^{\theta,+}_{\omega}(x)\, \chi_{(-\infty, -L)}(x)+ 
  \left( u_\omega^{\theta,+}, \;  
v_\omega^{\theta,+}, \; w_\omega^{\theta,+}\;
\right)\, \chi_{[-L, \infty)}(x),
\ene
and
\beq\label{2.12}
{\mathbf U}^{\theta, +}_{\omega,2}:= \left(
P.V. \frac1{\alpha(x)} + \frac{i \pi}{\alpha'(x_\omega)} \delta_D(x_\omega), 0, 0 \right) \chi_{(-L, \infty)}(x).
  \ene
  We have proven in \cite{jmpa} that,
\beq \label{2.16}
\left\|    \left( u_\omega^{\theta,+}, \;  
v_\omega^{\theta,+}, \; w_\omega^{\theta,+}\;
\right)\    \right\|_{L^2} \leq C,
\ene
where the constant $C$ is uniform for $ \omega$ in compact sets. This is the reason why
${\mathbf U}^{\theta, +}_{\omega, 1}$ is referred to as the regular part.
  Then,
  \beq \label{2.13}
P_\varphi(t)=  P_{\varphi, 1}(t) + P_{\varphi, 2}(t)
\ene
where, 
  \beq \label{2.14}
P_{\varphi, 1}(t):= \int_{\mathbb R }\, e^{-i \omega t}\, A(\omega) \left( {\mathbf U}^{\theta,+}_{\omega,1}, \varphi \right)d\omega ,
\ene
\beq \label{2.15}
P_{\varphi,2}(t):= \int_{\mathbb R }\, e^{-i \omega t}\, A(\omega) \left( {\mathbf U}^{\theta,+}_{\omega,2}, \varphi \right)d\omega .
\ene
 Then, using also \eqref{2.6} we obtain that 
$ \left( {\mathbf U}^{\theta,+}_{\omega,1}, \varphi \right)$ is  bounded for $ \omega \in [\omega_0,\omega_1]$ and since $A(\omega)$ is integrable, it follows from \eqref{2.9} that,

\beq\label{2.17}
\lim_{t \rightarrow \pm \infty }\, P_{\varphi, 1}(t)=0.
\ene
We now consider $P_{\varphi,2}(t)$ where we have to take into account the singularity of the solution.

 For  $ \varepsilon >0$ let  $ \rho_\omega(\pm \varepsilon)$ satisfy: $ \alpha (\rho_\omega
 (\pm \varepsilon)) = \pm \varepsilon$. Note that if $ \varepsilon$ is small enough,  $\forall \omega \in [\omega_0,\omega_1]$,   $\rho_\omega(\pm \varepsilon)$   is unique and there is a $\rho_0 >0$ such that, $ \left|\rho_\omega(\pm \varepsilon)\right| \geq \rho_0$.  Moreover,  $ \alpha(x)$ is one to one for $x \in[\rho_\omega(\varepsilon), \rho_\omega(-\varepsilon)]$ and  $\left|\frac{1}{\alpha\sp{\prime}(x)}\right|$ is uniformly bounded for $ \omega \,\in \,[\omega_0, \omega_1], \,x \in [\rho_\omega( \varepsilon), \rho_\omega(-\varepsilon)]$. Observe that this holds because  $N_e$ is  continuously differentiable and $\rho_\omega(\varepsilon)< \rho_\omega(x_\omega)<
 \rho_\omega(-\varepsilon) $ and $\alpha'(x_\omega)<0$.
 
 We decompose $P_{\varphi,2}(t)$ in the following way
\beq \label{2.18}
P_{\varphi,2}(t)=  P^{(1)}_{\varphi,2}(t)+ P^{(2)}_{\varphi,2}(t) ,
\ene
with,
 \begin{equation}\label{2.19}
  P^{(1)}_{\varphi,2}(t):=   \int 
   \left[  \int  \left(
   \frac{1}{\alpha (x)}   \left( \chi_{(-L, \rho_\omega(\eta ))}(x) +\chi_{(\rho_\omega(- \eta), \infty)}(x) \right)  \varphi(x) +  \frac{i \pi}{\alpha'(x_\omega)}\,  \varphi(x_\omega) \right)  dx  
   \right]    e^{-i \omega t}  A(\omega ) d\omega  ,
 \end{equation}
with $ \eta > \varepsilon$, and
\begin{equation}\label{2.20}
 P^{(2)}_{\varphi,2}(t):= \int e^{-i \omega t} \, A(\omega )\,  h(\omega)  d\omega  ,\,\textrm{where} \quad   h(\omega):= \lim_{\varepsilon \downarrow 0}\left[  \int_{\rho_\omega(\eta)}^{\rho_\omega(\varepsilon)}  \frac{1}{\alpha (x)}   \varphi(x) 
  dx  +   \int_{\rho_\omega(-\varepsilon)}^{\rho_\omega(-\eta)}   \frac{1}{\alpha (x)}   \varphi(x)  dx    \right].
\end{equation}
 
 Since the function inside the brackets in the right-hand side of \eqref{2.19} is uniformly bounded for  $\omega \in [\omega_0, \omega_1]$  and  $A(\omega)$  is integrable,  it follows from  \eqref{2.9} that,
 \begin{equation}\label{2.21}
\lim_{t \rightarrow \pm \infty}  P^{(1)}_{\varphi,2}(t)= 0.
 \end{equation}

Let $ \beta_\omega(z)$ be the inverse function to $ \alpha_\omega(x)$ i.e., $ \beta_\omega(\alpha_\omega (x))= x$, $ x \in [\rho_\omega(\eta), \rho_\omega(-\eta)]$. Then, changing the variable of integration into $ z:= \alpha_\omega(x)$ we obtain that the function $h(\omega)$ in \eqref{2.20} is equal to 
\begin{eqnarray*} 
h(\omega) &=& - \lim_{\varepsilon \downarrow 0}\left[  \int_{-\eta}^{-\varepsilon}  \frac{1}{z}  
\frac{\varphi(\beta_\omega(z)) }{\alpha'(\beta_\omega(z))}\;   dz  +   \int_{\varepsilon}^{\eta}   \frac{1}{z} \frac{\varphi (\beta_\omega(z)) }{\alpha'(\beta_\omega(z))}  \;  dz    \right]\\
&=& - \lim_{\varepsilon \downarrow 0}\left[  \int_{-\eta}^{-\varepsilon}   \frac{1}{z}  \;
\left(\frac{\varphi(\beta_\omega(z)) }{\alpha'(\beta_\omega(z))}-\frac{1}{\alpha'(\beta_\omega(0))} \,\varphi(\beta_\omega(0)) \right) 
\, dz   +   \int_{\varepsilon}^{\eta} 
  \frac{1}{z}\left( \frac{\varphi(\beta_\omega(z)) }{\alpha'(\beta_\omega(z))}  -\frac{1}{\alpha'(\beta_\omega(0))} \,\varphi(\beta_\omega(0)) \right)
\, dz   \right],
\end{eqnarray*}
and then, 
\beq\label{2.22}
h(\omega ) = - \int_{-\eta}^\eta     \frac{1}{z} \left( \frac{1}{\alpha'(\beta_\omega(z))} \,\varphi(\beta_\omega (z)) -
\frac{1}{\alpha'(\beta_\omega(0))} \,\varphi(\beta_\omega(0))\right)\, dz .
\ene
 Hence, $h (\omega)$ is bounded for $ \omega \in [\omega_0,\omega_1]$ and  as $A(\omega)$ is integrable, by \eqref{2.9},
  \begin{equation}\label{2.23}
 \lim_{t \rightarrow \pm \infty}\,  P^{(2)}_{\varphi,2}(t) = 0.
 \end{equation}
Then, we have proven,
\begin{theorem}
For every continuously differentiable $\varphi$ with compact support,
\beq \label{2.24}
 \lim_{t \rightarrow \pm \infty}\,  P_{\varphi}(t) = 0.
 \ene
\end{theorem}
 \noindent {\it Proof:} The theorem follows from \eqref{2.13}, \eqref{2.17}, \eqref{2.18}, \eqref{2.21} and \eqref{2.23}.

 Equation \eqref{2.24}, with  appropriate functions $\varphi$, implies that that as $ t \rightarrow \pm \infty$  the   part of the wave packet   in $[-R,-L]$ tends to zero for every $ R>0$. Since there is no heating for $ x \leq -L$ this means that the reflected part of the wave packet actually travels to $-\infty$ in the $x$ direction.  Furthermore, the wave packet tends to zero, as $\ t \rightarrow \pm \infty$  for  $ x \geq L$. Since we  assume that our problem is coercive (non propagative)  for $ x \geq H$ the wave packet cannot travel to plus infinite in the $x$ direction. {This implies that for large times  the transmitted part of the wave packet is dissipated by collisions.}
 
 Note that the condition  $ \omega _0> c |\theta|$ is only  used in order that for $ x \leq -L$ we have an incoming and a reflected plane wave, see \eqref{2.6}. Equation \eqref{2.24} also holds for   $ \omega_0 \leq c |\theta|$ with a similar proof. We omit the details.

\end{document}